\documentclass[aps,prb,twocolumn,shortbibliography,superscriptaddress,article]{revtex4-2}
\usepackage{amsmath}
\usepackage{amssymb}
\usepackage{graphicx}
\usepackage{array}
\usepackage{dcolumn}
\usepackage{subfigure}
\usepackage{color}
\usepackage{float}
\usepackage{xr-hyper}
\usepackage[hidelinks=true]{hyperref}
\hypersetup{
  colorlinks   = true, %Colours links instead of ugly boxes
  urlcolor     = blue, %Colour for external hyperlinks
  linkcolor    = blue, %Colour of internal links
  citecolor   = red %Colour of citations
}

\usepackage{epsfig}
\usepackage{epstopdf}
\usepackage{amsmath}
\usepackage{amsfonts}
\usepackage{amssymb}
\usepackage{hyperref}
\usepackage{bm}
\usepackage{makecell}
\usepackage{rotating}
\usepackage{hyperref}
\usepackage{multirow}
\usepackage{graphicx}
\usepackage{array} 
\usepackage{tabularx} % make sure this is in your preamble
\newcolumntype{Y}{>{\centering\arraybackslash}X}

\usepackage{chemformula} 
\usepackage[T1]{fontenc} 
\usepackage[percent]{overpic}
\usepackage{academicons} 
\usepackage{xcolor}
\usepackage{ragged2e}

\usepackage{graphicx}% Include figure files
\usepackage{dcolumn}% Align table columns on the decimal point
\usepackage{bm}% bold math
\usepackage{color}
\usepackage{comment}
\usepackage[hidelinks=true]{hyperref}
\usepackage{tikz,xcolor,hyperref}
\DeclareMathAlphabet\mathbfcal{OMS}{cmsy}{b}{n}

\usepackage{physics}

\definecolor{lime}{HTML}{A6CE39}

\usepackage{orcidlink}

\begin{document}

\title{Bond Disproportionation, Ligand Holes, and Persistent Spin Textures in Ag$_2$BiO$_3$}

\author{Atanu Paul}
\affiliation{Smart Ferroic Materials Center, Physics Department and Institute for Nanoscience and Engineering, University of Arkansas, Fayetteville, Arkansas 72701, USA}

\author{Subhadeep Bandyopadhay\orcidlink{0009-0007-8680-6310}}
\affiliation{Department of Physics, University of Duisburg-Essen, Lotharstrasse 1, 47057 Duisburg, Germany}

\author{Anupam Mondal\orcidlink{0009-0001-2672-7684}}
\affiliation{School of Physical Sciences, Indian Association for the Cultivation of Science, 2A \& 2B Raja S. C. Mullick Road, Jadavpur, Kolkata 700032, West Bengal, India}

\author{Indra  Dasgupta\orcidlink{0000-0002-6012-9530}}
\email{sspid@iacs.res.in}
\affiliation{School of Physical Sciences, Indian Association for the Cultivation of Science, 2A \& 2B Raja S. C. Mullick Road, Jadavpur, Kolkata 700032, West Bengal, India}

\date{\today}

\begin{abstract}
The origin of the proposed bond disproportionated insulating state of the non-centrosymmetric ($Pnn2$) phase of   
Ag$_{2}$BiO$_{3}$ is explored using first principles electronic structure calculations. The novel insulating state is elucidated by first considering the initially proposed centosymmetric metallic ($Pnna$) phase of Ag$_{2}$BiO$_{3}$. Our calculations reveal that the valence skipping Bi$^{4+}$ ions
in this phase are better described as Bi$^{3+}\underline{L}$ with completely filled 
Bi-(6$s$) states and a ligand hole.  However, phonon calculations indicate that the metallic ($Pnna$) state is dynamically unstable. Structural stability is achieved through breathing distortions of the oxygen octahedra, resulting in two inequivalent Bi sites and a reduction of symmetry to the $Pnn2$ phase. Electronic structure calculations further reveal that the $Pnn2$ phase is a bond 
disproportionated insulator where the nominal charge state of Bi is described by : 
2[Bi$^{3+}\underline{L}$ (Bi$^{4+}$)] $\rightarrow$  Bi$^{3+}\underline{L}^{2-\delta}$ (Bi1$^{5+}$) + Bi$^{3+}\underline{L}^{\delta}$ (Bi2$^{3+}$),
highlighting the crucial role of ligand holes in driving the insulating state. Next we have investigated the 
 electronic structure of  Ag$_{2}$BiO$_{3}$ in the insulating ($Pnn2$) phase including spin-orbit coupling. Our density functional theory (DFT ) calculations complemented by \textbf{k.p} model Hamiltonian analysis reveal persistent spin-textures around the $X$ and $Y$ high symmetry points of the orthorhombic Brillouin zone imposed by non-symmorphic symmetry, positioning 
  Ag$_{2}$BiO$_{3}$ as a promising  candidate for spintronic applications.
\end{abstract}

\maketitle

\section{Introduction}
%%%%%%%%%%%%%%%%%%%%%%%%%%%%%%%%%%%%%%%%%%%%%%%%%%%%%%
Bismuthates have long held a special status in condensed matter research due to the wide range of remarkable quantum phenomena they exhibit, including novel and intriguing metal-insulator transitions \cite{P1_38,A_Paul_2019,BFO_MI,z5tm-cvgn, PhysRevMaterials.8.044204, S_Bandyopadhyay_2024}, enhanced thermoelectric properties~\cite{BiTe_Thermo, BiTe_thermo1, Bi_Thermo, pnas.1608794113}, nontrivial topological phases~\cite{Bi2Te3_topological, PhysRevLett.122.167401, PhysRevLett.111.136802, PhysRevLett.108.117403, science.1189924, science.1245085, pnas.1900527116} and superconductivity~\cite{Changdar2025, Shen2017}. Among them, BaBiO$_{3}$ is one of the most extensively studied compounds, known for its bond-disproportionated insulating state, which upon hole doping evolves into a superconducting state~\cite{BBO_Super,SLEIGHT2015152}. At low temperatures, BaBiO$_{3}$ is insulating and exhibits a characteristic distortion from the ideal cubic perovskite structure. In particular, the oxygen octahedra surrounding the Bi ions undergo breathing distortions along all three cubic crystallographic directions.

In recent years, its monovalent chemical analogue Ag$_{2}$BiO$_{3}$ has attracted considerable attention. Although Ag$_{2}$BiO$_{3}$ exhibits an insulating phase similar to that of BaBiO$_{3}$, the insulating phase is non-centrosymmetric and structurally distinct, featuring a combination of corner- and edge-sharing BiO$_{6}$ octahedra. Ag$_{2}$BiO$_{3}$ was first reported in
$Pnna$ phase~\cite{jansen1} with all bismuth atoms occupying the same
crystallographic site and this left with the only possibility
for assigning the nominal charge state of bismuth to be 4+. Bi in 4+
charge state can only be associated with metallic conductivity due to half filled 6$s^1$ orbital of bismuth
and this uncompensated spin will possibly lead to paramagnetism. However,
the samples were found to be insulating and diamagnetic. In
order to settle the inconsistency with respect to the valence state
of bismuth in Ag$_{2}$BiO$_{3}$, the crystal structure was reinvestigated by employing temperature
dependent X-ray diffraction and neutron powder diffraction~\cite{jansen2}. These investigations
revealed that the crystallographic phase to be $Pnn2$, in
which there exist two distinct crystallographic sites for bismuth, allowing for a possible charge ordering as to be Bi$^{3+}$ and
Bi$^{5+}$ in agreement with the fact that Bi$^{4+}$ is valence skipping. Further cooling of the sample lowers the symmetry
, resulting in the $Pn$ phase where the charge ordering phenomena
remain unaffected. \\ An earlier first principles electronic structure
calculation on Ag$_{2}$BiO$_{3}$ using generalized gradient approximation (GGA) within the FP-LAPW method as well as  ultrasoft pseudopotential plane wave method recovered the
insulating   $Pnn2$ phase~\cite{liu},  but the origin of
charge ordering induced insulating state  remained largely unexplored. A recent report confirms that the $Pnn2$ phase of  Ag$_{2}$BiO$_{3}$ to be a bond-disproportionated insulator featuring Weyl nodal chains~\cite{AGO_bond}. The non-centrosymmetric insulating $Pnn2$ state is predicted to be ferroelectric and to host Rashba-Dresselhaus spin-orbit coupling (SOC), leading to characteristic spin textures at specific high-symmetry points of the Brillouin zone (BZ)~\cite{Nature,S_Bandyopadhyay_2021}. Furthermore, it has been proposed that if the undisproportionated semi-metallic phase ($Pnna$) with equivalent Bi site could be stabilized, Ag$_{2}$BiO$_{3}$ may host topological phases, including a hour glass nodal net semimetal, Dirac semimetal and hourglass Dirac semimetal states~\cite{Singh, Fu}.

In the present paper we have investigated the electronic structure of Ag$_{2}$BiO$_{3}$
in both $Pnna$ and $Pnn2$ phases and clarified the origin of
metal-insulator transition using various indicators of chemical bonding. Lattice dynamics calculations are carried out in the $Pnna$
phase in order to identify the unstable phonon modes. Our calculations of the phonon modes at the $\Gamma$ point are in agreement with a previous report  ~\cite{Nature} and the condensation of these modes lead to polar non-centrosymmetric $Pnn2$ and $Pn$ phases. 
Finally, we have studied the electronic structure of the $Pnn2$ phase including spin-orbit coupling. An excursion through
the Brillouin zone of the non-centrosymmetric $Pnn2$ phase from one high-symmetry point to the other reveal 
nontrivial spin splitting of the bands and consequent diverse spin
textures induced by spin-orbit coupling. In particular, we have identified non-symmorphic symmetry enforced persistent spin textures near the $X$ and $Y$ high symmetry points. Unlike the previously reported Dresselhaus\cite{Nature,S_Bandyopadhyay_2021}  spin-texture in Ag$_2$BiO$_3$, the spin orientations near $X$ and $Y$ points are constrained predominantly along fixed crystallographic directions by the glide symmetries of the \textit{Pnn2} structure \cite{Tao2018,LL_Tao}. This establishes the coexistence of qualitatively distinct spin-orbit coupling regimes within the same Brillouin zone and highlights Ag$_2$BiO$_3$ as a promising platform for symmetry controlled spin transport.   
The remainder of the paper is organized as follows. In section 2 we have discussed the structure of Ag$_{2}$BiO$_{3}$ and computational
details, section 3 is devoted to our results and discussions.
Finally, conclusions are given in section 4.
%%%%%%%%%%%%%%%%%%%%%%%%%%%%%%%%%%%%%%%%%%%%%%%%%%%%%%%%%%%%%%%%%%%%%%%%%%

\section{Structure and Computational details}

\begin{figure*}%[t]
\centering
\includegraphics[width=0.95\linewidth]{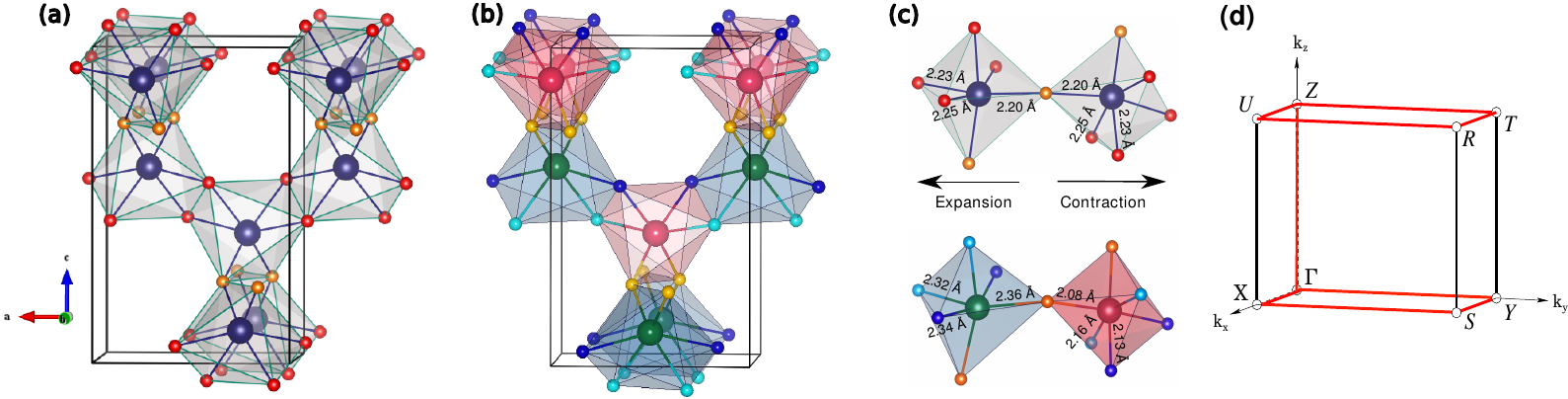}
\caption{Crystal structure of (a) the $Pnna$ and (b) the $Pnn2$ phases. Only the BiO$_6$ octahedral network is shown, where Bi and O atoms are represented by larger and smaller spheres, respectively. (c) Two neighboring BiO$_6$ octahedra with equal volumes in the $Pnna$ phase (top) become inequivalent, forming contracted and expanded octahedra in the $Pnn2$ phase (bottom). This symmetry-lowering transition also leads to the emergence of more inequivalent oxygen sites in the $Pnn2$ phase. (d) Irreducible part of the orthorhombic Brillouin zone for space group Pnn2 (34) along with high symmetric k-points. The solid red lines correspond to the high-symmetric k-path along which the band structures have been drawn.}
  \label{fig1}
\end{figure*}

The crystal structure of Ag$_{2}$BiO$_{3}$ in the $Pnna$ phase is shown
in Fig.~\ref{fig1} (a). The unit cell contain four formula units with 24
atoms. The structure of Ag$_{2}$BiO$_{3}$ can be
considered as one dimensional corrugated  chain along
$x$-axis, formed by edge shared BiO$_{6}$ octahedra and each
chains are connected via common corner shared oxygens
along $y$-axis and forming a three dimensional network. In
case of centrosymmetric  $Pnna$ phase all Bi atoms occupy the same crystallographic sites with distorted octahedral environment
offered by the oxygens. Later analysis revealed that a more reliable structure admits the space group $Pnn2$
with two distinct crystallographic sites for bismuth (Bi1 and Bi2) and thereby removes the inversion symmetry (see Fig.~\ref{fig1} (b)).
%%%%%%%%%%%%%%% Figure1: Figure of the Pnna and Pnn2 Structure

%\textbf { Atanu : Please modify the figure as suggested} \\
The difference in crystal structure between $Pnna$ and
$Pnn2$ phase, is manifested by the smaller Bi1O$_{6}$ 
and larger Bi2O$_{6}$ octahedra, in case of $Pnn2$ phase as
shown in Fig.~\ref{fig1} (b) and (c) and they are arranged alternately.
The Bi-O bond
length corresponding to each phases, are mentioned in
each octahedron. The lattice parameters and atomic positions were reported by Deibele and Martin~\cite{jansen1} for the \textit{Pnna} phase, and by Oberndorfer et al.~\cite{jansen2} for the \textit{Pnn2} phase.
The electronic structure calculations for the weakly correlated Ag$_{2}$BiO$_{3}$
are performed using density functional theory employing
local density approximation (LDA) with the projector
augmented-wave (PAW)~\cite{blochl} potentials as implemented in
the Vienna ab-initio simulation package (VASP)~\cite{vasp1,vasp2}. The
valence electrons are described by a plane wave basis set in
the calculation. The plane-wave
cut-off energy is set to  500 eV along with a 12$\times$12$\times$12 $\Gamma$
centered k-mesh to perform the k-space integration using
the tetrahedron method for four formula units. In our
calculation the valence states for Ag, Bi and O are taken
as 5$s^1$ 4$d^{10}$, 5$d^{10}$ 6$s^2$ 6$p^3$ and 2$s^2$ 2$p^4$ respectively. The self-
consistent calculations are considered to be converged only
when the energy convergence achieved is less than 10$^{-5}$
eV. Electronic structure calculations are carried out for the
experimental structure for both the phases~\cite{jansen1, jansen2}. Fig.1(d) displays the irreducible part of the orthorhombic Brillouin zone(Bz) of Ag$_2$BiO$_3$, along with various high-symmetry points, \textit{viz.}, $\Gamma(0,0,0)$, $X(1/2,0,0)$, $S(1/2,1/2,0)$, $Y(0,1/2,0)$, $Z(0,0,1/2)$, $U(1/2,0,1/2)$, $R(1/2,1/2,1/2)$, and $T(0,1/2,1/2)$. The phonon calculations are carried out
using the finite displacement method as implemented in the PHONOPY code~\cite{Phonopy}. A \(2 \times 2 \times 2\) supercell of the optimized centrosymmetric $Pnna$ structure of Ag$_2$BiO$_3$ is used to compute the phonon band structure. The structural relaxations are performed until the Hellmann–Feynman forces on all atoms are reduced below 0.005\,eV/\AA.

To analyze the electronic structure in real space, we employed the N-th order muffin-tin orbital (NMTO) downfolding method based on self-consistent linear muffin tin orbital method (LMTO) calculations~\cite{LMTO, LMTO1, NMTO}. In this approach, an effective low-energy Hamiltonian is constructed by integrating out high-energy degrees of freedom from the full LMTO Hamiltonian in an energy-selective manner. The resulting NMTO basis yields localized orbitals corresponding to the chosen energy window. For isolated bands, these orbitals span the same Hilbert space as the corresponding Wannier functions and thus provide a Wannier-like representation of the electronic structure. These orbitals were used to construct real-space Wannier function plot and to examine the bonding characteristics of the system.
%The crystal
%orbital Hamiltonian population (COHP)~\cite{dron} has been computed to analyze the chemical bonding, as implemented
%in the tight-binding linear muffin tin orbital (TB-%LMTO)~\cite{NMTO1,NMTO2,NMTO3}
%code in the atomic sphere approximation.\\
%%%%%%%%%%%%%%%%%%%%%%%%%%%%%%%%%%%%%%%%%%%%%%%%%%%%%%%%%%%%%%%%%%%%%%%%%%
\section{Results and discussions}
\subsection{Electronic structure of Ag$_{2}$BiO$_{3}$ for the $Pnna$ Phase}

\begin{figure}[h]
\centering
  \includegraphics[height=4.8cm]{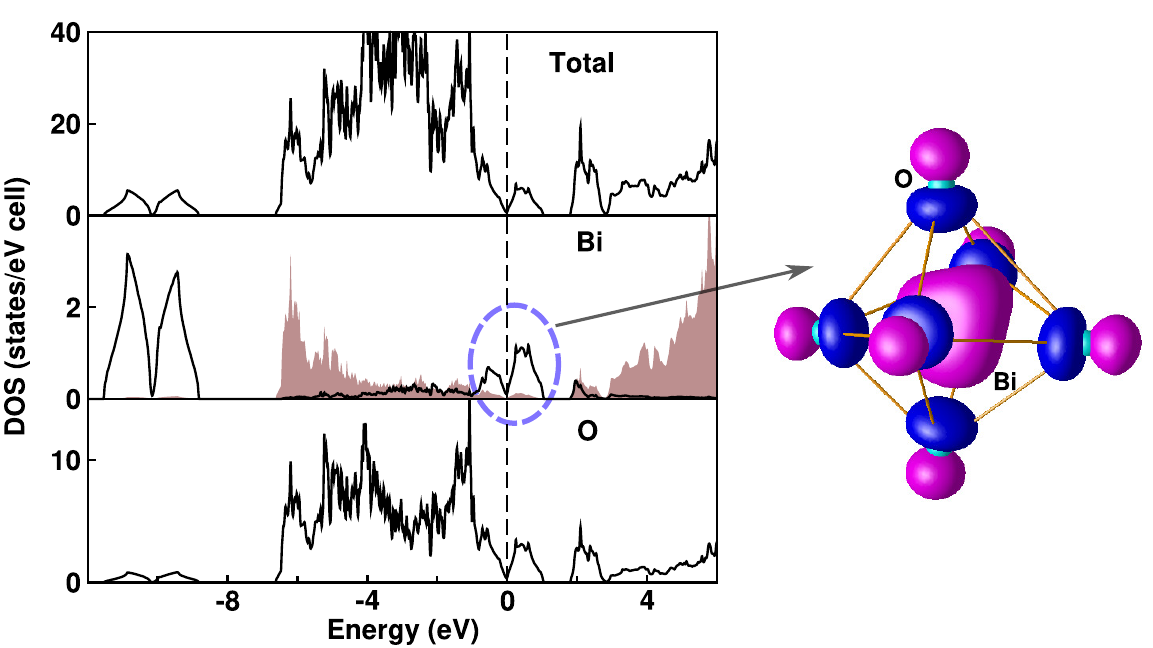}
  \caption{$Pnna$ phase: Total and partial densities of states (DOS) of Bi-$s$, Bi-$p$ (filled curve), and O-$p$ states. The Fermi level is set to zero. The inset shows the Bi-$s$ Wannier function corresponding to the states near the Fermi level, plotted within the BiO$_6$ octahedron.}
  \label{fig2}
\end{figure}

In order to understand the origin of the insulating state of Ag$_{2}$BiO$_{3}$, we first analysed the electronic structure of  the  $Pnna$ phase. We have plotted the total and partial
density of states (DOS) for the Bi-6$s$, Bi-6$p$ and O-$p$ of Ag$_2$BiO$_3$ in the  $Pnna$ phase in
Fig.~\ref{fig2}. The total DOS is metallic. It was expected that Bi should be in the 4+($s^1$)
electronic configuration and half filled Bi-$6s$ state will be situated around the Fermi level.
%%%%%%%%%%%%%%%%%%% Figure 2 DOS Pnna and  Wannier Function

Interestingly, the Bi-6$s$ states are seen to be more
pronounced much below the Fermi level between -12 eV to
-8 eV energy range, suggesting the charge state of Bi does not
fulfil our expectation. Rather the Bi-6$s$ and O-$p$ states
are strongly hybridized in the energy range -12 eV to 1
eV, forming a bonding state predominantly of Bi-$6s$ character around -10 eV region,
and antibonding
states around the Fermi level, mainly dominated by
O-$p$ states. The weight of O-$p$ density of states in the
half-filled antibonding interaction region implies that the
holes are situated in the ligands instead of Bi sites. The density of states in the energy range
-6 eV to 1 eV are dominated by O-$p$ states, whereas
the Bi-6$p$ states are found to be above the Fermi level ( See Fig.~\ref{fig2} ).

In order to corroborate the existence of the ligand hole 
we have plotted the Bi-6$s$ Wannier function using NMTO-downfolding method near the Fermi level (See  Fig. \ref{fig2} inset), we find strong hybridization of Bi-$s$ with O-$p$ states where the O-$p$ lobes are directed
towards the spherically symmetric Bi-6$s$ orbital, forming
a strong $sp\sigma$ type antibonds between Bi-6$s$ and
O-$p$ ($a_{1g}$) orbitals in the BiO$_{6}$
octahedral unit and the predominant presence of oxygen states provide credence to the
ligand hole character as can be seen in the plot. This type
of antibonding between Bi-6$s$ and O-$p$, has been already
reported for bismuth perovskites~\cite{sawatzky,P2_40}. The nominal valence of Bi in the $Pnna$ phase may be assigned as $Bi^{4+}$ $\rightarrow$ $Bi^{3+} \underline{L}$
This state is however not stable as it was suggested~\cite{verma} that Bi$^{4+}$ is valence skipper and likely to  be
stabilized by charge disproportionation forming Bi$^{3+}$ and Bi$^{5+}$. Therefore it is expected that there should be some symmetry
lowering structural transition in Ag$_{2}$BiO$_{3}$.\\
%%%%%%%%%%%%%%%%%%%%%%%%%%%%%%%%%%%%%%%%%%%%%%%%%%%%%%%%%%%%%%%%%%%%%%%%%%%%%%%%%%%%%%%%%%
\begin{figure}[h!]
\centering
\includegraphics[height=8.8cm, angle=270]{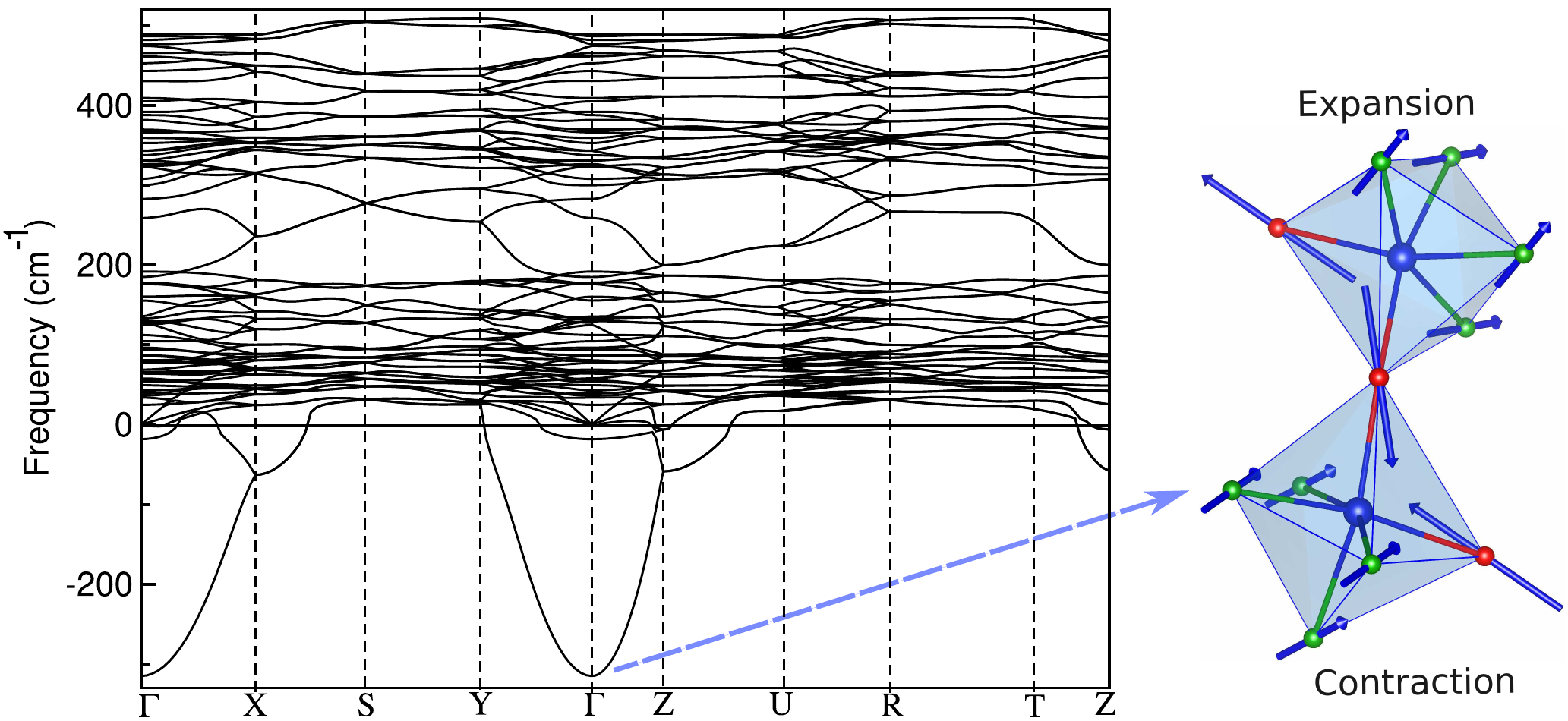}
\caption{Phonon dispersion of the $Pnna$ phase. Inset shows the eigenvectors corresponding to the soft mode at 315$i$ cm$^{-1}$ in consecutive octahedra. Only the oxygen's movement is shown.}
\label{Pnna_Phonon}
\end{figure}

\subsection{ Emergence of the $Pnn2$ phase }
Phonon band structure of the $Pnna$ phase (shown in Fig.\ref{Pnna_Phonon}) exhibits multiple instabilities both at the center ($\Gamma$-point) as well as at the boundaries of the Brilliouin zone. In particular, we find two unstable phonon modes at the $\Gamma$-point having frequencies of 315$i$ cm$^{-1}$ and 18$i$ cm$^{-1}$ respectively, consistent with previous studies\cite{Nature}. Larger unstable $\Gamma$-phonon mode corresponds predominantly to a breathing like distortion of the BiO$_6$ octahedral networks (with additional distortion of the Ag atoms), as shown in Fig.\ref{Pnna_Phonon}, and is  referred to as $\Gamma_2^-$  irreducible representation. Condensation of the $\Gamma_2^-$ atomic distortion lowers the symmetry to a $Pnn2$, producing two inequivalent Bi sites with different BiO$_6$ octahedral volumes. This is expected to lead to charge disproportionation at the Bi sites, leading to an insulating nature of the $Pnn2$ phase. Detailed analysis of the electronic properties of the $Pnn2$ phase is given in the following section. Moreover, simultaneous presence of edge and corner shared BiO$_6$ octahedra lifts the inversion symmetry while condensation of the $\Gamma_2^-$ distortion, making $Pnn2$ a polar structure. He $et~ al.$\cite{Nature} reported a electric polarization of 8.8 $\mu C/cm^2$ for the $Pnn2$ structure.  On the other hand, the smaller unstable $\Gamma$-phonon mode corresponds to a $\Gamma_3^-$ atomic distortion, which also lowers the symmetry to a polar $Pna2_1$ structure. However $Pna2_1$ has not been realized so far in experiment. 
Interestingly, when $\Gamma_2^-$ and $\Gamma_3^-$ are condensed together, symmetry of the system is further lowered to a monoclinic $Pn$ phase, which is realized experimentally at low temperature~\cite{OBERNDORFER_2006, Takagi_2021}. Our calculation finds that the polar $Pn$ phase is nearly degenerate in energy with the $Pnn2$ phase, but give rise to a larger band gap, in agreement with the report by He et. al.~\cite{Nature}. We note that, despite overall consistency, our calculated phonon band structure differs at several zone-boundary points when compared with the literature \cite{Nature}. Specifically,  along $S-Y$ and $U-R$ paths, due to absence of phonon instabilities at $Y$ and $U$-point respectively. These differences most likely arise from the choice of exchange–correlation functional~\cite{Compare_phonon_1,Compare_phonon_2}. In the present work, we employ the LDA, whereas  He et al.~\cite{Nature} have used PBEsol in their calculations.\\
Next we have analyzed the stability of the $Pnn2$ phase.
The phonon dispersion of the $Pnn2$ phase calculated within the LDA is found to exhibit small imaginary phonon modes (within approximately -13 cm$^{-1}$) along the X-S, S-Y, U-R, R-T, and T-Z directions, while no imaginary mode is observed at the $\Gamma$ point (See Appendix). Since these weak instabilities may originate from the overbinding tendency of LDA, we further examined the volume dependence of the phonon spectrum by uniformly expanding the relaxed lattice parameters by $1\%$, corresponding to an approximately $3\%$ increase in the unit-cell volume. With this volume expansion, all phonon frequencies become positive throughout the Brillouin zone, indicating that the observed weak instability is strongly volume dependent and does not represent an intrinsic dynamical instability of the experimentally observed $Pnn2$ phase(See Appendix).

\begin{figure}[h!]
\centering
\includegraphics[height=6.0cm]{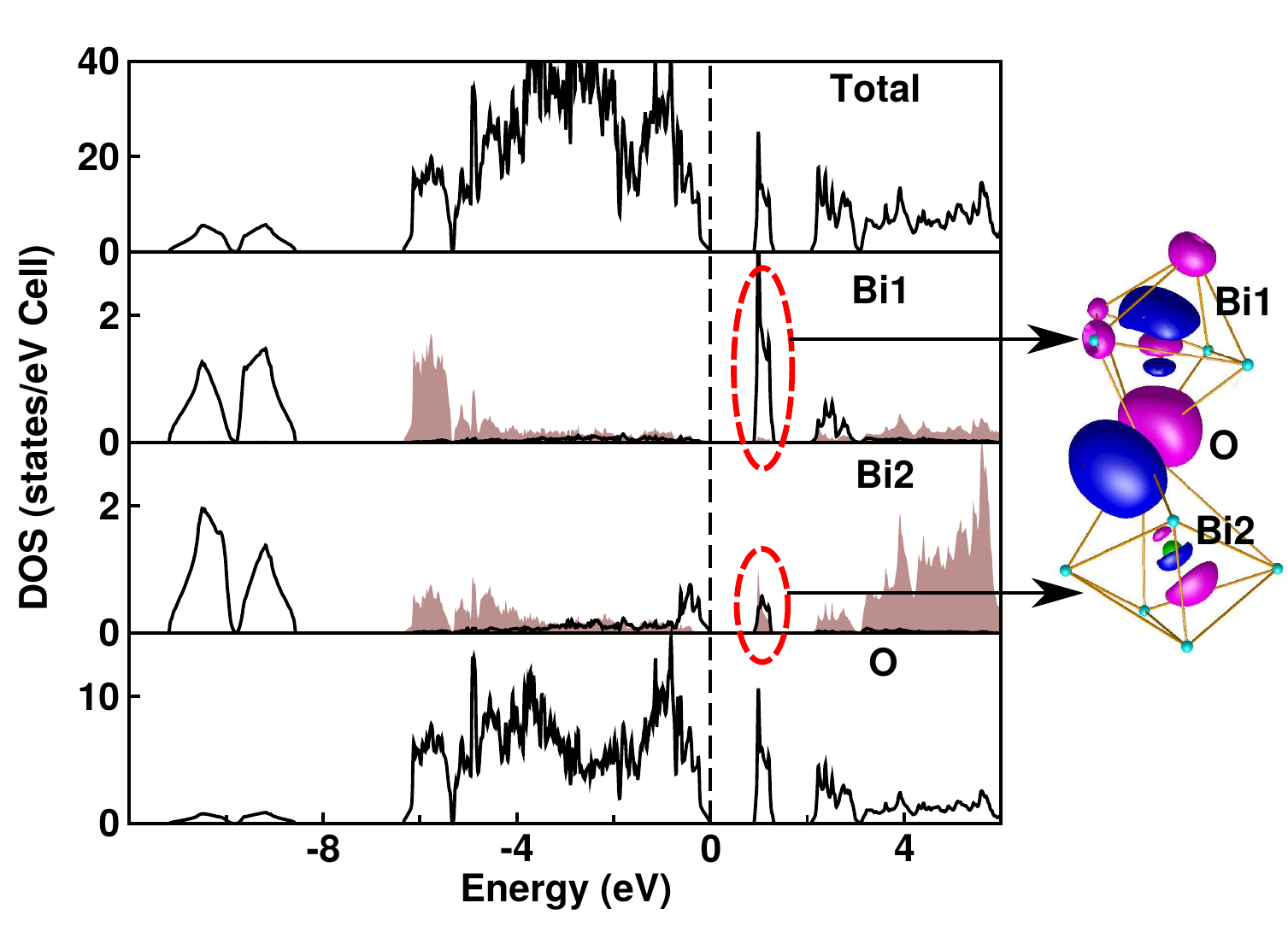}
\caption{Total and partial DOS of Bi1-$s$, Bi1-$p$, Bi2-$s$, Bi2-$p$ and O-$p$ for the $Pnn2$
phase. The filled curves corresponds to Bi-$p$. The Fermi level is set at zero. Inset is the plot of O-$p$ Wannier function.}
\label{fig4}
\end{figure}

\subsection { Electronic Structure of the $Pnn2$ phase}
Next we have investigated the electronic structure of the 
 $Pnn2$ phase with two inequivalent Bi sites, Bi1 and Bi2 belonging to the smaller and larger octahedra respectively. The total DOS
and partial DOS of Bi1-6$s$, Bi1-6$p$,
Bi2-6$s$, Bi2-6$p$ and O-$p$ in the $Pnn2$ phase are shown
in Fig.~\ref{fig4}. It gives a semiconducting solution within LDA  and the
 band gap is calculated to be 0.69 eV in very good agreement to 
the reported value of band gap  for Ag$_{2}$BiO$_{3}$~\cite{jansen1}. It may be noted that the band structure calculated within GGA is very similar to LDA except the band gap is slightly increased to 0.72 eV. Similar to the $Pnna$ phase Bi1-6$s$ and Bi2-6$s$ states are occupied and lies in between -12 eV to -8 eV energy range. There is strong 
hybridization between Bi-6$s$ and O-$p$ states in the $Pnn2$ phase. In
fact, the hybridization between Bi1-6$s$ and O-$p$ states
are stronger than Bi2-6$s$ and O-$p$ states due to
shorter Bi1-O bond length in the smaller octahedra.  The  $sp\sigma$ anti-bonding state  Bi-6$s$-O$p$ due to Bi-6s and O-p hybridization,
is now situated above the Fermi level, indicating  that the ligand hole
is mainly associated with the smaller Bi1O$_{6}$ octahedral unit. So the hole pairs condense spatially onto the collapsed Bi1O$_{6}$ octahedron, while the oxygens corresponding to the expanded octahedron Bi2O$_{6}$ remain largely occupied.

The above arguments are further justified by calculating the O-$p$
Wannier function around 1 eV above the Fermi level (see Fig.~\ref{fig4} inset ). The plot of the Wannier function reveal that the unoccupied state has dominant contribution from the oxygens residing at the collapsed octahedra
(Bi1O$_{6}$), confirming that the ligand hole  belongs to the Bi1O$_6$ octahedra.  Our calculation point to the fact that the nominal charge state of Bi in the expanded octahedron is Bi$^{3+}$ {$\underline{L}^{\delta}$} while for the collapsed octahedron it is Bi$^{3+} \underline{L}^{2-\delta}$ as Bi$^{5+}$ has an energetically deep 6$s$ shell  which strongly suppresses Bi$^{3+}$ - Bi$^{5+}$
charge disproportionation and rather leads to bond  disproportionation~\cite{Hase}.  The core level photoelectron spectroscopy of Bi 4$f$ in the $Pnn2$ phase exhibits two peaks separated by a binding energy difference of approximately 0.5 eV \cite{liu} suggesting that the two different Bi states are due to the two distinct Bi atoms in its unit cell where the nominal charge state of both are close to the value of Bi$^{3+}$, supporting the ligand hole scenario as revealed in our calculations. %%%%%%%%%%%%%%%%%%%%%%%%%% 

 %%%%%%%%%%%%%%%%%%%%%%%%%%%%%%%%%%%%%%
In conclusion,  the transformation from  metallic $Pnna$ to insulating $Pnn2$ is accompanied by bond disproportionation where  the charge
state of Bi in the $Pnn2$ phase may be assigned as 2$s^2$$\underline{L}$$^1$ $\rightarrow$
$s$$^2 \underline{L}^{\delta}$ + $s$$^2$$\underline{L}$$^{2-\delta}$.\\

%%%%%%%%%%%%%%%%%%%%%%%%%%%%%%%%%%%%%%%%%%%%%
\subsection{Spin-orbit induced band splitting and persistent spin textures}
\begin{figure*}[t!]
\centering
\includegraphics[width=1\linewidth]{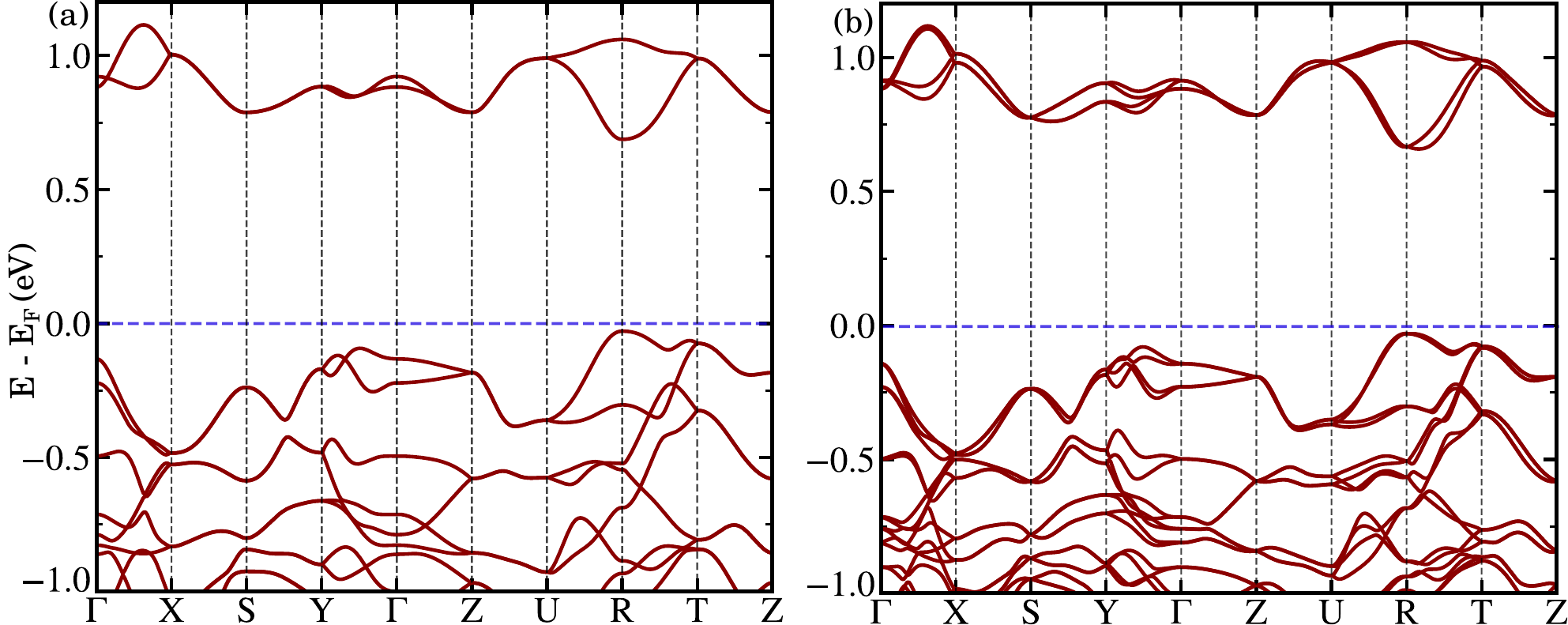}
\caption{Band structure of Ag$_2$BiO$_3$ calculated (a) without spin--orbit coupling and 
(b) with spin--orbit coupling in the vicinity of the Fermi energy.}
%\textcolor{red}{Bands are plotted w.r.t to the $E_f$, why then $E_f$ is shown around 1 eV?}
\label{band_structure}
\end{figure*}
In this section, we have explored the role of spin orbit coupling in the $Pnn2$ phase. As mentioned earlier, the non-centrosymmetric $Pnn2$ phase is predicted to be ferroelectric.  In ferroelectric systems, broken inversion symmetry produces spatial gradient of electrostatic potential that gives rise to an effective momentum-dependent magnetic field, $\mathbf{B}(\mathbf{k})$, in the  rest frame of an electron. This field $\mathbf{B}(\mathbf{k})$  couples with the electron spin through spin--orbit coupling \cite{DiSante2013GeTeRashba,S_Bandyopadhyay_2020}. As a result, the spin degeneracy of the electronic bands can be lifted even in a nonmagnetic system. The corresponding SOC Hamiltonian may be written as
$H_{\mathrm{SOC}}=\mathbf{B}(\mathbf{k})\cdot\boldsymbol{\sigma},$
where $\boldsymbol{\sigma}$ represents the Pauli spin matrices. The form of $\mathbf{B}(\mathbf{k})$ is dictated by the symmetry of the crystal and may contain both linear and higher-order momentum-dependent terms \cite{Zunger,WANG2026116090,Wang_book}. This effective field locks the spin direction to the crystal momentum, producing nontrivial spin textures in reciprocal space. For a Bloch state $u_{n\mathbf{k}}$, the spin texture is defined as the expectation value
$\langle \mathbf{S}_n(\mathbf{k})\rangle=
\langle u_{n\mathbf{k}}|\hat{\mathbf{S}}|u_{n\mathbf{k}}\rangle,
$
where $n$ denotes the band index.
Depending on the underlying symmetry, the resulting spin texture may take Rashba, Dresselhaus, persistent, radial, or more complex forms ~\cite{K_Dutta_2023,K_Dutta_PRM}. \\ Although crystallographic point-group symmetry provides the first guideline for classifying possible spin textures, it is not always sufficient by itself. The spin texture near a particular high-symmetry point is governed more directly by the little group of that wave vector, together with the orbital character of the relevant electronic states\cite{Zunger,S_Bandyopadhyay_2021}. In this respect, the importance of non-symmorphic symmetries has been argued for the existence of persistent ST \cite{Tao2018}. Consequently, different high-symmetry points within the same material can host distinct spin textures, even though the global crystal symmetry remains unchanged.

It has been reported that the conduction band of Ag$_2$BiO$_3$ at $R$ point in presence of SOC displays  Dresselhaus type spin splitting with characteristic spin texture \cite{Nature}. Similarly it has been shown Bi-$s$ + O-$p_z$ character of the $Z$-point of Ag$_2$BiO$_3$ with $C_{2v}$ point group symmetry promotes Dresselhaus spin texture~\cite{S_Bandyopadhyay_2021}. 
In the following we shall calculate the spin-texture at the unexplored  $X$ and $Y$ points of the orthorhombic BZ of Ag$_2$BiO$_3$ where at the $X$ and $Y$ points the impact of the non-symmorphic symmetry is expected to be important. \\ The non-spin polarized band structure of Ag$_2$Bi$O_3$ is shown in Fig. \ref{band_structure}(a). As argued earlier, here O-$p_z$ states admixed with Bi-$s$ states form the lowest conduction band, whereas the valence bands are primarily dominated by O-$p$ states.
The bandstructure including SOC has been shown in Fig. \ref{band_structure} (b). Splitting of the valence and conduction bands can be seen in the perpendicular direction($U$-$R$-$T$) of the polarization axis whereas
there is no splitting in the parallel direction ($\Gamma$-$Z$). This suggests the possibility of Rashba-Dresselhaus splitting in the system. 
%At 'R' point (k=\frac{1}{2},\frac{1}{2},\frac{1}{2}) bandstructure shows direct band gap, so splitting around R is important as
%application point of view. We will concentrate on the splitting around R point in this paper.
The conduction band displays relatively large spin-orbit induced splitting in comparison to the valence band. 
\subsubsection{The Symmetry analysis}
Ag$_2$BiO$_3$ crystallizes in $Pnn2$ space group (number: 34), which is a non-symmorphic space group with broken inversion symmetry. The point group symmetry is $C_{2v}$. The system contains the following non-symmorphic symmetry operations: (a) glide reflection $\overline{M}_x$, which consists of a mirror reflection about the $x$ = 0 plane along with a translation of $(\frac{1}{2},\frac{1}{2},\frac{1}{2})$: 
\begin{eqnarray}
    \overline{M}_x :(x,y,z) \rightarrow (-x + \frac{1}{2} , y + \frac{1}{2} , z +\frac{1}{2}),
\end{eqnarray} (b) glide reflection $\overline{M}_y$, which consists of mirror reflection about the y=0 plane $M_y$ followed by the $(\frac{1}{2},\frac{1}{2},\frac{1}{2})$ translation: 
\begin{eqnarray}
    \overline{M}_y :(x,y,z) \rightarrow (x + \frac{1}{2} , -y +\frac{1}{2} , z +\frac{1}{2}).
\end{eqnarray}
There are no screw rotations present in the system.  In the following, we shall discuss the spin-textures around high symmetry X and Y points in terms of a low energy $\mathbf{k}\cdot\mathbf{p}$ model Hamiltonian and compare our results with DFT calculations. In particular, we shall show that  the conduction band  of Ag$_2$BiO$_3$ at X and Y points display persistent spin texture imposed by non-symmorphic symmetry similar to BiInO$_3$ \cite{Tao2018}. PST is the property to maintain an unidirectional spin orientation in the momentum space, that are robust against the spin-independent disorder leading to large spin lifetime which is crucial for spintronics.

% It ends here. 
\subsubsection{X point}
\begin{figure}[h!]
    \centering
    \includegraphics[width=1\linewidth]{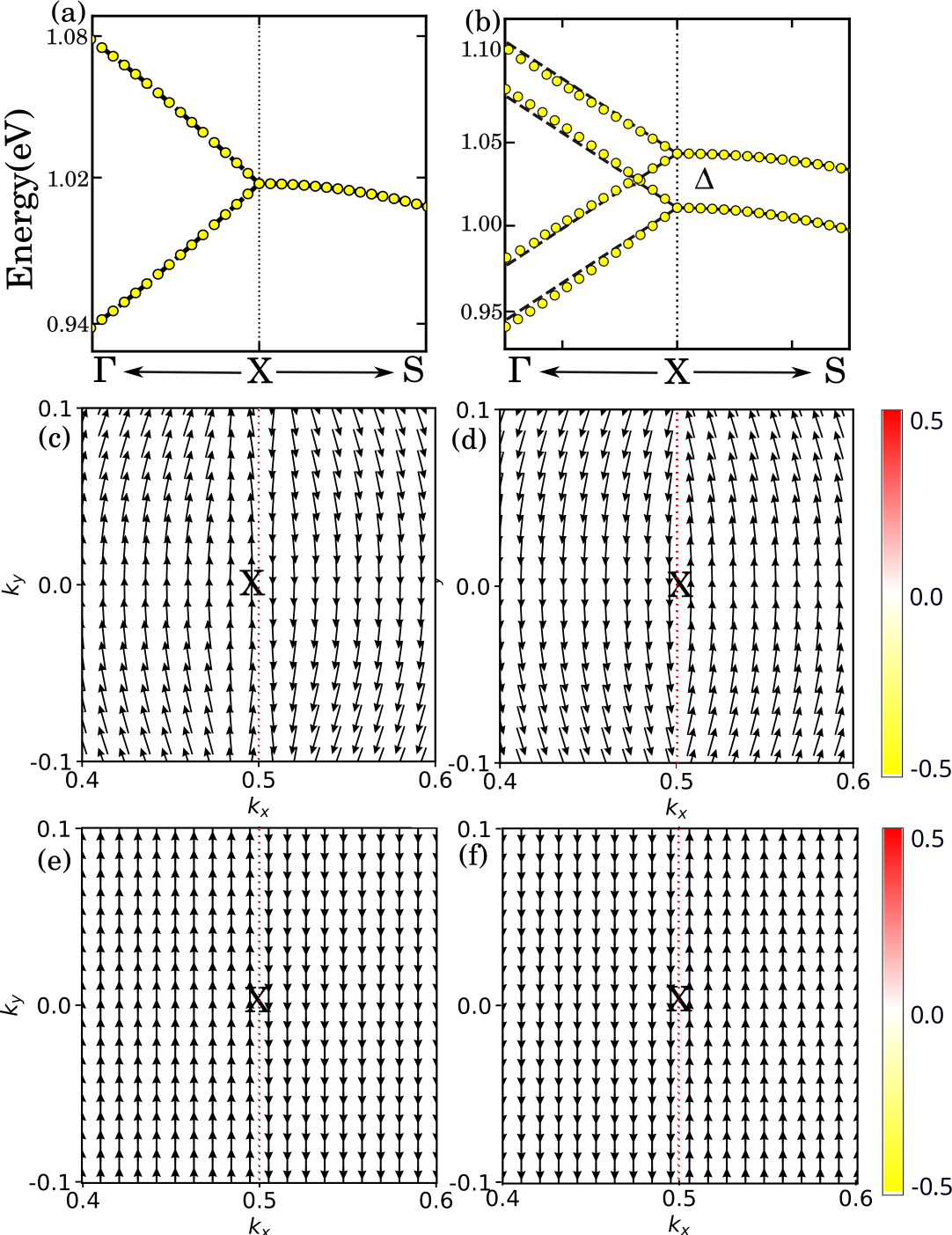}
    \caption{Band structure and spin texture (ST) of Ag$_2$BiO$_3$ in the $k_x$-$k_y$ plane. (a,b) Band structures without and with SOC along the path $\Gamma(0,0,0)\rightarrow X(\frac{1}{2},0,0)\rightarrow S(\frac{1}{2},\frac{1}{2},0)$. The SOC-split bands are fitted using the four-dimensional $\mathbf{k}\cdot\mathbf{p}$ model Hamiltonian. In (b) the yellow circles correspond to DFT band structure whereas the black dashed lines are the fitted bands  using our model Hamiltonian. We fitted the DFT bands over a small segment of the  $\Gamma(0,0,0)\rightarrow X\left(\frac{1}{2},0,0\right)\rightarrow S\left(\frac{1}{2},\frac{1}{2},0\right)$ path, spanning reciprocal coordinates from $(0.42,0,0)$ to $(0.5,0.075,0)$. The reciprocal coordinates are given in reduced reciprocal lattice units,i.e.,$\mathbf{k}=k_1\mathbf{b}_1+k_2\mathbf{b}_2+k_3\mathbf{b}_3$.
    (c,d) Inner and outer conduction-band spin-texture branches around the $X$ point. The spin expectation values are predominantly locked along the $s_y$ direction, indicating persistent spin texture, while the $s_z$ component remains negligible throughout the plotted region, as shown by the color bar. (e) and (f) corresponds to the spin textures obtained from our low energy $\mathbf{k}\cdot\mathbf{p}$ model which are in agreement with the DFT results.}
    \label{fig:X}
\end{figure}

%\textcolor{red}{We may cite https://www.nature.com/articles/s41467-018-05137-0 while discussing the PST's around X and Y. }
The DFT band structures around the X point along the path $\Gamma \rightarrow X \rightarrow S$ in a narrow k-range without and with spin-orbit coupling are shown by circles in Fig. \ref{fig:X} (a) and (b) respectively. Along the path $X \rightarrow \Gamma$ the four-fold degenerate band without spin-orbit coupling at the X point splits into a pair of  doublets. Inclusion of spin orbit coupling,  further splits them into singlets along the  $X\rightarrow\Gamma$ path. However along the path $X\rightarrow S$, the bands without SOC are constrained to remain fourfold degenerate and inclusion of SOC splits them into a pair of doublets. The DFT spin-textures are displayed in Fig. \ref{fig:X} (c) and (d) show unidirectional spin orientation in the momentum space characteristic of persistent spin texture. \\
We first discuss the origin of the band splitting in the absence of spin--orbit coupling along the $\Gamma \rightarrow X$ path. Along this  path, a generic $k$-point has the form $(k_x,0,0)$. On this path, the little group contains the symmetry operations $\overline{M}_y$ and $\Theta=T\overline{M}_x$, where $T$ is the time-reversal symmetry operator. Since the Hamiltonian commutes with the anti-unitary operator $\Theta=T\overline{M}_x$, Kramer's theorem can be used to determine whether this symmetry enforces a degeneracy. If $\Theta^2 = -1$, the anti-unitary symmetry imposes a Kramer's-like degeneracy, whereas if $\Theta^2 = 1$, no such degeneracy by symmetry is enforced. Along the path $\Gamma\rightarrow X$, we find $\Theta^2 = T^2 \overline{M}_x^2 = e^{-ik_y - ik_z} = 1$. Therefore, the anti-unitary operator does not impose a degeneracy along $\Gamma \rightarrow X$, leading to a pair of doublets [see Fig. \ref{fig:X} (a)].  At the same time, $\overline{M}_y$ is a symmetry operator  along this path, so the Bloch states can be chosen as eigenstates of $\overline{M}_y$. Since $\overline{M}_y^2=1$ at the $X$ point, the corresponding eigenvalues are $\pm 1$, and we can write $\overline{M}_y\psi_X^{\pm}=\pm\psi_X^{\pm}.$ Thus, the bands along $\Gamma \rightarrow X$ can belong to different $\overline{M}_y$ sectors. Since these sectors are not required to be degenerate, the bands are allowed to split along this direction.\\ We now consider the $X \rightarrow S$ path, where a generic $k$-point has the form $(\frac{1}{2},k_y,0)$. Along this path, the little co-group contains $\overline{M}_x$ and the anti-unitary symmetry $\Theta=T\overline{M}_y$. For this path, we obtain $\Theta^2 = T^2 \overline{M}_y^2 = e^{-i\pi} =-1$. Therefore, by the same Kramer's-theorem argument, the anti-unitary symmetry $\Theta=T\overline{M}_y$ enforces a twofold band degeneracy along the $X\rightarrow S$ path. This explains the behavior observed in the band structure without SOC in Fig. \ref{fig:X}(a) where the bands are split along $\Gamma \rightarrow X$, as no anti-unitary degeneracy is enforced, whereas they remain degenerate along $X\rightarrow S$ due to the non-symmorphic anti-unitary symmetry.
\\\\ Inclusion of SOC at the X point  splits the fourfold degenerate states into two doublets with splitting $\Delta$. In the presence of SOC, the glide operators act on both, the sublattice and spin degrees of freedom. The sublattice degrees of freedom emerges due to the nontrivial Bloch phase generated by the fractional translation in non-symmorphic space group operations. This nontrivial momentum dependent Bloch phase modify the little group algebra at the Brillouin zone boundary\cite{LL_Tao}.  Let us first consider the $\Gamma \rightarrow X$ path, where a generic $k$-point has the form $(k_x,0,0)$ and the little group contains $\overline{M}_y$ and the anti-unitary symmetry $\Theta=T\overline{M}_x$. For a spin-$1/2$ system, $T^2=-1$. Therefore, along this path we obtain
$\Theta^2 = T^2\overline{M}_x^2 = (-1)\times e^{-ik_y - ik_z} \times (-1) = 1 .$ Thus, although the Hamiltonian commutes with the anti-unitary operation $\Theta=T\overline{M}_x$, this symmetry does not enforce a Kramer's-like degeneracy along $\Gamma \rightarrow X$. As a result, the bands are allowed to be non-degenerate along this path, and the twofold degeneracies are now lifted [see Fig. \ref{fig:X} (b)].
We now consider the $X \rightarrow S$ path, where a generic $k$-point $(\frac{1}{2},k_y,0)$ lies on the Brillouin-zone boundary and the little group contains $\overline{M}_x$ and the anti-unitary symmetry $\Theta=T\overline{M}_y$. In this case, the square of the anti-unitary operator becomes
$\Theta^2 = T^2 \overline{M}_y^2 = (-1) \times e^{-i\pi} \times (-1) = -1 .$ Therefore, the anti-unitary symmetry $\Theta=T\overline{M}_y$ enforces a twofold degeneracy between $\psi_k$ and $\Theta \psi_k$ along the $X-S$ path. Hence, even in the presence of SOC, the bands remain doubly degenerate along $X \rightarrow S$ path. 
\\\\ To describe the persistent spin-texture around high symmetry point X we first construct the $\mathbf{k}\cdot\mathbf{p}$ model Hamiltonian using the non-symmorphic symmetry operations.   Previous work\cite{Tao2018,LL_Tao} has shown that non-symmorphic orthorhombic space groups with broken inversion symmetry can be classified according to their fractional translation vectors, and  these symmetry operations impose constraints on the effective spin-orbit field near high symmetry points, leading to a symmetry enforced persistent spin texture. Since our system belongs to one of the space groups in this classification, we construct the effective $\mathbf{k}\cdot\mathbf{p}$ Hamiltonian following the same symmetry-guided framework.
 We define the small momentum measured from $X$ as $\mathbf{q}=(q_x,q_y,0),$ where $k_x=\frac{1}{2}+q_x,\qquad k_y=q_y,\qquad k_z=0.$ The four states are described using both spin and sublattice degrees of freedom. The spin degree of freedom is represented by the Pauli matrices $\sigma_i$ and identity matrix $\sigma_0$, while the sublattice degree of freedom is represented by the Pauli matrices $\tau_i$ and identity matrix $\tau_0$. Consequently, the allowed terms in the effective Hamiltonian can be written as products $\tau_i \otimes \sigma_j$, constrained by the non-symmorphic glide symmetries and time-reversal symmetry. In the chosen basis, the relevant non-symmorphic symmetry generators at $X$ are represented as
$\overline{M}_x=i\tau_z\otimes\sigma_x, \qquad \overline{M}_y=\tau_y\otimes\sigma_y,$ and the time-reversal operator is
$T=\tau_0K\otimes i\sigma_y,$ where $K$ denotes complex conjugation. Imposing invariance under $\overline{M}_x$, $\overline{M}_y$, and $T$, the most general Hamiltonian up to linear order in $\mathbf{q}$ can be written as
\begin{eqnarray}
H_X(\mathbf{q})&=&\epsilon_X(\mathbf{q})\tau_0 \otimes \sigma_0+\zeta \tau_y \otimes \sigma_y+\mathrm{a} q_x \tau_0\otimes \sigma_y \notag \\
&+&\mathrm{b} q_y \tau_0\otimes \sigma_x+\notag\xi_1 q_x \tau_y\otimes \sigma_0+\xi_2 q_x \tau_x\otimes \sigma_x\\
&+&\xi_3 q_x \tau_z\otimes\sigma_z+\xi_4 q_y \tau_x\otimes \sigma_y 
\end{eqnarray}
Here, $\epsilon_X(\mathbf{q})=\mathrm{E}_X+\mathrm{A}_xq_x^2+\mathrm{A}_yq_y^2$ describes the spin-independent dispersion. The term $H_0=\zeta \tau_y\otimes \sigma_y$ is momentum independent and is responsible for splitting the fourfold degenerate state at $X$ into two doublets. Since $\overline{M}_y=\tau_y\otimes \sigma_y,$ this term can be written as $H_0=\zeta \overline{M}_y.$ Therefore, the two doublets are labeled by the eigenvalues of $\overline{M}_y=\pm1$ and are separated by
$\Delta=2\zeta.$ The remaining momentum-dependent terms determine the dispersion away from the $X$ point and the Hamiltonian projected within each $\overline{M}_y$-labeled doublet reduces, to first order, to
\begin{eqnarray}
H_X^{\pm}(q_x)=\pm\zeta\sigma_0+ \mathrm{a}^{\pm}q_x\sigma_y ,
\end{eqnarray}
where $a^{\pm}=\sqrt{(a\pm\xi_1)^2+(\xi_2\mp\xi_3)^2}.$
The corresponding eigenvalues are $E^{\pm}_{\eta}(q_x)=\pm\zeta+\eta a^{\pm}q_x,\qquad\eta=\pm1.$
Thus, the model naturally explains the DFT band structure: SOC first splits the fourfold degenerate state at $X$ into two doublets, and the linear-in-$q_x$ term further splits each doublet into two branches along the $X$-$\Gamma$ direction.
The same projected Hamiltonian also explains the persistent spin texture. Since the effective Hamiltonian within each doublet contains only the $\sigma_y$ spin matrix to leading order, the spin expectation value is constrained to lie along the $y$ direction:
$\langle \sigma_y\rangle\neq0,\qquad\langle \sigma_x\rangle=0,\qquad\langle \sigma_z\rangle=0.$
Therefore, the spin direction does not rotate with momentum around the $X$ point, giving rise to a persistent spin texture. The fitted $\mathbf{k}\cdot\mathbf{p}$ bands are shown by the black dashed lines in Fig.~\ref{fig:X}(a) and (b). The fitting gives
$\zeta=0.016\ \mathrm{eV},$
corresponding to a doublet separation
$\Delta=2\zeta=32\ \mathrm{meV}.$ The extracted effective spin-orbit parameters are
$\mathrm{a}^{+}\simeq \mathrm{a}^{-}=0.429\ \mathrm{eV}$\AA. The $\mathbf{k}\cdot \mathbf{p}$ spin textures are displayed in Fig. \ref{fig:X}(e) and (f). The results obtained from the model Hamiltonian agree well with the DFT calculations around neighbourhood of the $X$ point where linear in q term approximation is valid. At larger momenta, higher-order terms become increasingly important and  generate a finite $x$ component ($S_x$) of the spin expectation value, leading to a gradual deviation from the ideal persistent spin texture.
\subsubsection{$Y$ point }
\begin{figure}[h!]
   \centering
    \includegraphics[width=1\linewidth]{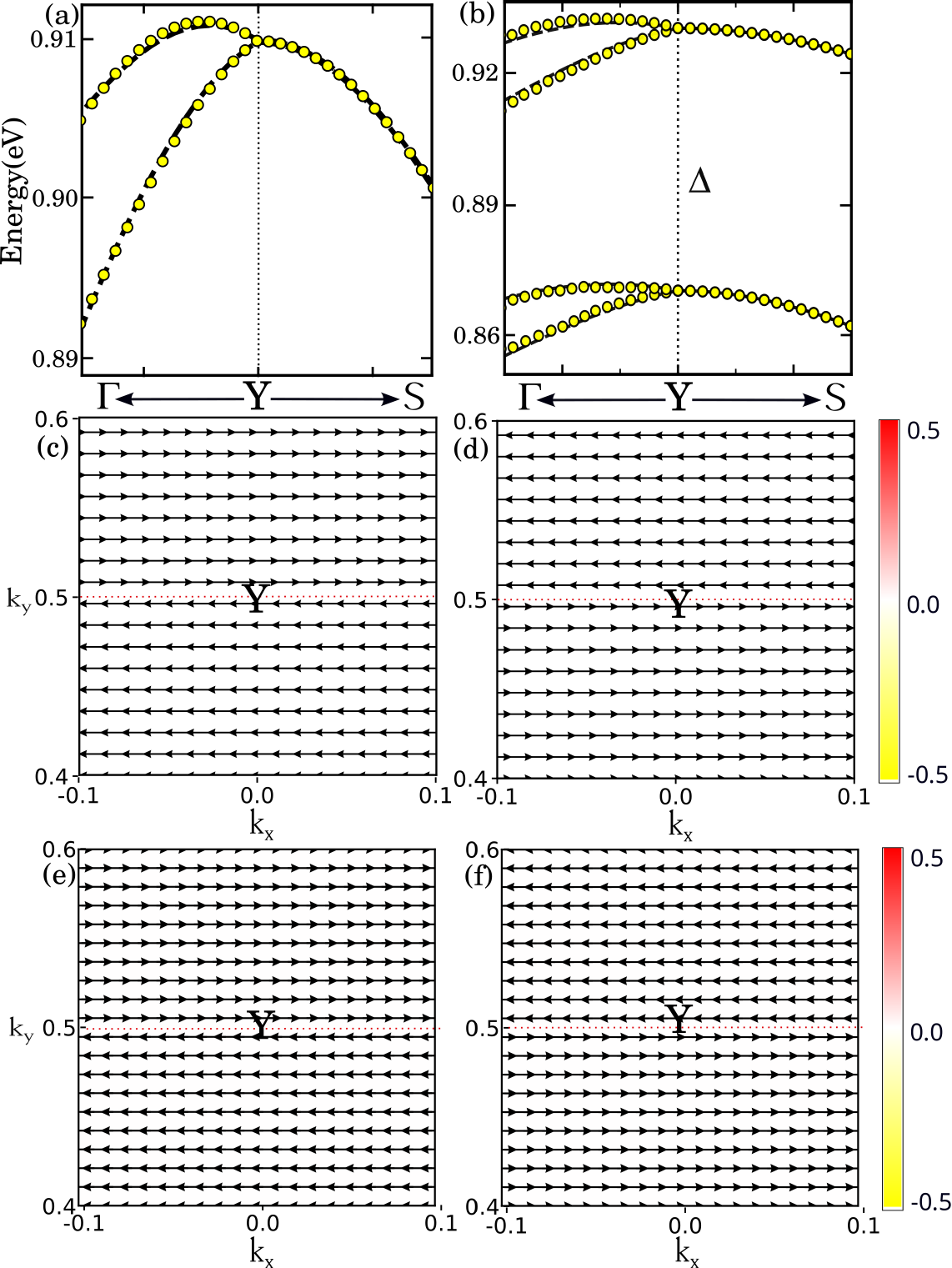}
    \caption{ Band structure and spin texture (ST) of Ag$_2$BiO$_3$ in the $k_x$-$k_y$ plane. (a,b) Band structures without and with SOC along the path $\Gamma(0,0,0)\rightarrow Y(0,\frac{1}{2},0)\rightarrow S(\frac{1}{2},\frac{1}{2},0)$. The SOC-split bands are fitted using the four-dimensional $\mathbf{k}\cdot\mathbf{p}$ model Hamiltonian. The yellow circles correspond to the DFT band structure whereas the black dashed lines correspond to the fitted band structure using our $\mathbf{k}\cdot\mathbf{p}$ model Hamiltonian.  We fitted the DFT bands over a small segment of the  $\Gamma(0,0,0)\rightarrow Y\left(0,\frac{1}{2},0\right)\rightarrow S\left(\frac{1}{2},\frac{1}{2},0\right)$ path, spanning reciprocal coordinates from $(0,0.42,0)$ to $(0.075,0.5,0.0)$ in reciprocal lattice units. (c,d) Inner and outer conduction-band spin-texture branches around the $Y$ point. The spin expectation values are predominantly locked along the $s_x$ direction, indicating persistent spin texture, while the $s_z$ component remains negligible throughout the plotted region, as shown by the color bar. (e) and (f) corresponds to the spin textures we got from our low energy $\mathbf{k}\cdot\mathbf{p}$ model which are in agreement with the DFT results.}
    \label{fig:Y}
\end{figure}

The DFT band structure around the Y point along the path $\Gamma \rightarrow Y \rightarrow S$ without and with spin-orbit coupling is shown by circles in Fig. \ref{fig:Y} (a) and (b) respectively. The band structure is very similar to the X point and the spin-texture shown in Fig.\ref{fig:Y}(c) and \ref{fig:Y}(d) displays persistent nature now parallel along the $k_x$ direction. \\ 
A similar analysis applies to the Y point. Exactly at $Y$, with $ \mathbf{k}_Y=\left(0,\frac{1}{2},0\right),$
the little co-group symmetry  is $C_{2v}$. 
The $\Gamma \rightarrow Y \rightarrow S$ path can be understood in the same way by interchanging the roles of $x$ and $y$. Without SOC, along $\Gamma \rightarrow Y$, a generic $k$-point is $(0,k_y,0)$, and the little group contains $\overline{M}_x$ and $\Theta=T\overline{M}_y$. Here $\Theta^2=T^2\overline{M}_y^2=e^{-ik_x - ik_z}=1$, so the anti-unitary symmetry does not enforce degeneracy. Since $\overline{M}_x$ is a symmetry operator along this path, the bands can be classified by different $\overline{M}_x$ sectors and are allowed to split. Along $Y \rightarrow S$, a generic $k$-point is $(k_x,\frac{1}{2},0)$, and the relevant anti-unitary symmetry is $\Theta=T\overline{M}_x$. In this case, $\Theta^2=T^2\overline{M}_x^2=e^{-i\pi}=-1$, which enforces a twofold degeneracy along the $Y-S$ line.

 The same symmetry argument holds upon the inclusion of SOC where  the glide operators now act on both sublattice and spin space. Along $\Gamma \rightarrow Y$, $\Theta=T\overline{M}_y$ gives $\Theta^2=(-1)\times e^{-ik_x}\times(-1)=1$, so the bands can be non-degenerate. At the $Y$ point, SOC splits the fourfold manifold into two Kramer's doublets separated by $\Delta$, distinguished by the spinful $\overline{M}_x$ glide sectors. Along $Y \rightarrow S$, $\Theta=T\overline{M}_x$ gives $\Theta^2=(-1)\times e^{-i\pi}\times(-1)=-1$, and therefore the twofold degeneracy between $\psi_k$ and $\Theta\psi_k$ is protected.
The $\mathbf{k}\cdot\mathbf{p}$ model Hamiltonian around the $Y$ point can be constructed in a completely analogous way to the $X$-point Hamiltonian, with the roles of $x$ and $y$ interchanged. Around the $Y$ point, the relevant symmetry generators are $\overline{M}_x$, $\overline{M}_y$, and the time-reversal symmetry operator $T=\tau_0K\otimes i\sigma_y$, where $K$ is the complex conjugation operator. 
Here we are interested in the $k_x,k_y$ plane, and $\mathbf{k}$ is measured with respect to the $Y$ point. This Hamiltonian is the $Y$-point counterpart of the already discussed Hamiltonian for the $X$-point.  In first-order perturbation theory, the perturbation does not mix the two $\overline{M}_x$ doublets. Therefore, the effective Hamiltonian within each doublet can be written as
\begin{eqnarray}
    H^{\pm}(q_y)=\pm\zeta\sigma_0+\mathrm{b}^{\pm}q_y\sigma_x .
\end{eqnarray}
 Thus, near the $Y$ point, the effective spin-orbit field inside each doublet is proportional to $\sigma_x$. Therefore, the spin is forced by symmetry to lie along the $x$ direction. The $\mathbf{k}\cdot\mathbf{p}$ band structure is shown with the dashed lines in Fig. \ref{fig:Y}(a) and (b) while the spin textures are displayed in Fig. \ref{fig:Y} (e) and (f).    Values of fitted parameters are $\Delta = 2\zeta = 71$ meV, $\mathrm{b}^{+} = -0.12$ eV \AA  and $\mathrm{b}^{-} = -0.074$ eV \AA.  The low-energy $\mathbf{k}\cdot\mathbf{p}$ model accurately describes the DFT results within a small neighbourhood of the Y point where linear in q terms dominate the $\mathbf{k}\cdot\mathbf{p}$ Hamiltonian.
\section{Conclusion}
Using first principles electronic structure calculations we have clarified the origin of the proposed bond disproportionated insulating state in the non-centrosymmetric ($Pnn2$ ) phase of Ag$_2$BiO$_3$. We first consider the initially proposed $Pnna$ phase of Ag$_2$BiO$_3$ where all Bi atoms occupy the same crystallographic sites. Our calculations reveal that the valence skipping Bi$^{4+}$ ions in this phase are better described as Bi$^{3+}\underline{L}$, with completely filled Bi 6-s states and a ligand hole. Our phonon calculations further  reveal  that the semi-metallic ($Pnna$ ) phase is dynamically unstable. Structural stability is achieved through a breathing distortion of oxygen octahedra, leading to two in-equivalent Bi sites and the reduction of the symmetry to the $Pnn2$ phase. The resulting insulating  phase is not  only non-centrosymmetric but also structurally distinct featuring a combination of corner and edge shared BiO$_6$ octahedra. Electronic structure calculations reveal that the $Pnn2$ phase is bond disproportionated insulator where the nominal charge state of Bi is described by: $2[\mathrm{Bi}^{3+}\underline{L} (\mathrm{Bi}^{4+})] \rightarrow \mathrm{Bi}^{3+}\underline{L}^{2-\delta} (\mathrm{Bi}1^{5+}) + \mathrm{Bi}^{3+}\underline{L}^{\delta} (\mathrm{Bi}2^{3+})$. The energetically deep 6s shell of Bi$^{5+}$ strongly suppresses $\mathrm{Bi}^{3+} - \mathrm{Bi}^{5+} $ charge disproportionation. The bond disproportionated insulating state is also supported by core-level photo emission spectroscopy, highlighting the importance of the ligand holes in understanding the insulating state of Ag$_2$BiO$_3$. \\ Finally we have investigated the electronic structure of the insulating $Pnn2$ phase of Ag$_2$BiO$_3$ including spin-orbit coupling. Our DFT calculations complemented with symmetry adapted low energy $\mathbf{k}\cdot\mathbf{p}$ model confirm the presence of persistent spin textures at the high symmetric X and Y points of the orthorhombic Brillouin zone imposed by non-symmorphic symmetry. The realization of the persistent spin textures at the high-symmetry X and Y points in addition to Dresselhaus spin textures at the R and Z points reported earlier underscore the importance of Ag$_2$BiO$_3$ in spintronic applications. 
\section*{Data availability}
The data supporting the findings of this study, including the input files, calculated band structures, spin-texture data are available from the corresponding author upon reasonable request.
\section*{Acknowledgments}
ID thanks Prof. D D Sarma for sharing his valuable insights on the metal-insulator transition, and this article is dedicated to the occasion of his 70-th birthday. ID also thanks Prof. O. K. Andersen and Prof T. Saha-Dasgupta for discussions related to this work.
 AM acknowledges CSIR, India (Grant No.09/0080(17786)/2024-EMR-I) for the research fellowship. \\

\section*{Appendix}
\begin{figure}[h!]
    \centering
    \includegraphics[width=0.95\linewidth]{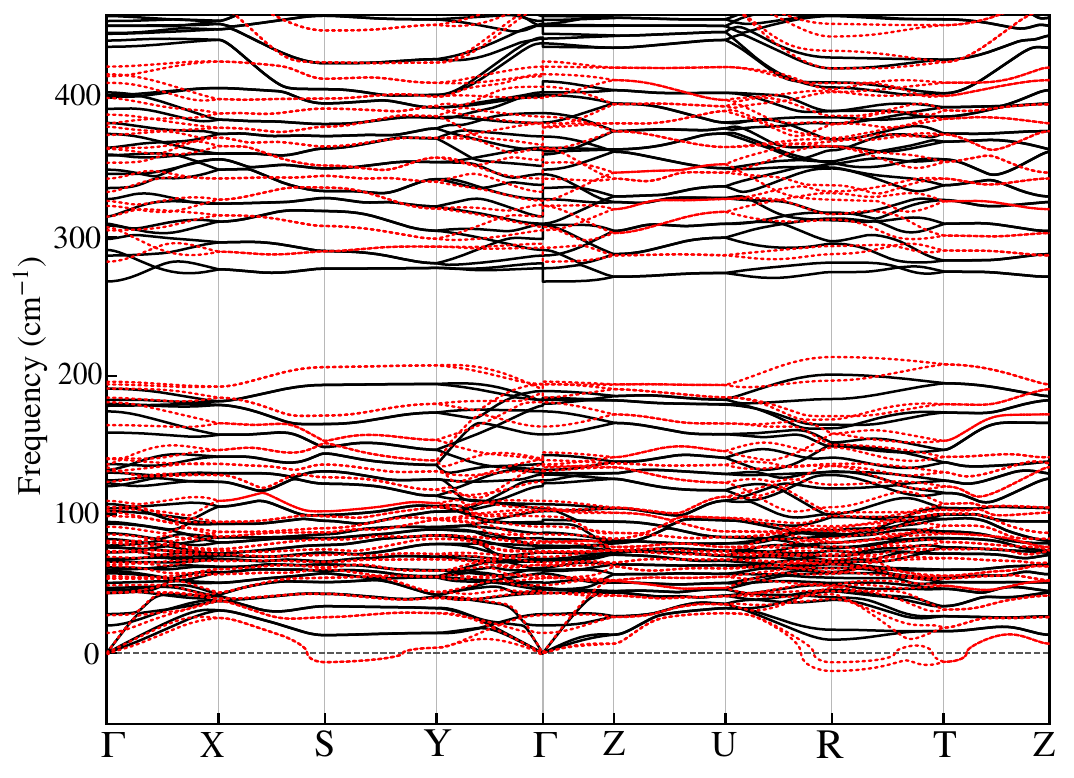}
    \caption{LDA-calculated phonon dispersions of the \textit{Pnn2} phase at the relaxed equilibrium volume (red dotted lines) and with a $3\%$ increase in the unit-cell volume (black solid lines).}
    \label{fig:Ag2BiO3_phonon_LDA_Pnn2}
\end{figure}

Fig.~\ref{fig:Ag2BiO3_phonon_LDA_Pnn2} shows the phonon dispersion of the \textit{Pnn2} phase at the LDA-relaxed equilibrium volume and after a $3\%$ increase in the unit-cell volume. The disappearance of the weak imaginary modes upon volume expansion demonstrates the strong sensitivity of the low-frequency phonon modes to the lattice volume.

%%%REFERENCES%%%
\bibliography{manuscript} %You need to replace "rsc" on this line with the name of your .bib file
\end{document}